\documentclass{article}
\usepackage{spconf,amsmath,epsfig}

\usepackage[]{graphicx}
\graphicspath{{Figures/}}
\usepackage{caption}
\usepackage{subfig}
\usepackage{amsmath}
\usepackage{amssymb}
\usepackage{epsfig}
\usepackage{cite}
\usepackage{color}
\usepackage{balance}
\usepackage{amsmath}
\usepackage{amsfonts} 
\usepackage{graphicx}
\usepackage{pdfpages}
\usepackage[T1]{fontenc} 
\usepackage{array}
\usepackage{url}

\usepackage{color}
\usepackage{wrapfig}
\usepackage{lipsum}
\usepackage[hidelinks]{hyperref}
\usepackage{multirow}
\usepackage{tabularx}
\usepackage{gensymb} 
\usepackage{breqn}
\usepackage{array}
\usepackage{booktabs}
\usepackage{gensymb}

\title{Impact of COVID19-induced Lockdown on Air Quality in Ireland}

\name{Dewansh~Kaloni$^{1}$, Yee~Hui~Lee$^{2}$, and Soumyabrata Dev$^{1,3}$
\thanks{The ADAPT Centre for Digital Content Technology is funded under the SFI Research Centres Programme (Grant 13/RC/2106) and is co-funded under the European Regional Development Fund.}
\thanks{Send correspondence to S.\ Dev: \url{soumyabrata.dev@ucd.ie}}
}
\address{
	$^{1}$~The ADAPT SFI Research Centre, Dublin, Ireland \\
	$^{2}$~School of Electrical and Electronic Engineering, Nanyang Technological University, Singapore\\
	$^{3}$~School of Computer Science, University College Dublin, Dublin, Ireland
}

\begin{document}
%
\maketitle
\begin{abstract}
Air pollution has been a long-existing problem for most of the major metropolitan cities of the world. Several measures including strict climate laws and reduction in the number of vehicles were implemented by several nations. However, in the recent wake of the COVID19 pandemic, there has been a renewed interest in revisiting the problem of low air quality. Several countries implemented strict lockdown measures halting the vehicular traffic and other economic activities, in order to reduce the spread of COVID19. In this paper, we analyze the impact of such COVID19-induced lockdown on the air quality of the atmosphere. Our case study is based in the city of Dublin, Ireland. We analyze the average concentration of common gaseous pollutant majorly responsible for industrial and vehicular pollution, \textit{viz.} nitrogen dioxide (NO$_2$). These concentrations are obtained from the tropospheric column of the atmosphere collected by Sentinel-5P, which is an earth observation satellite of European Space Agency. We observe that Dublin had a significant drop in the level of NO$_2$ concentration, owing to the strict lockdown measures implemented across the nation.
\end{abstract}


\begin{figure*}
\centering
\subfloat[24-Mar-2020]{\includegraphics[height=0.2\textwidth]{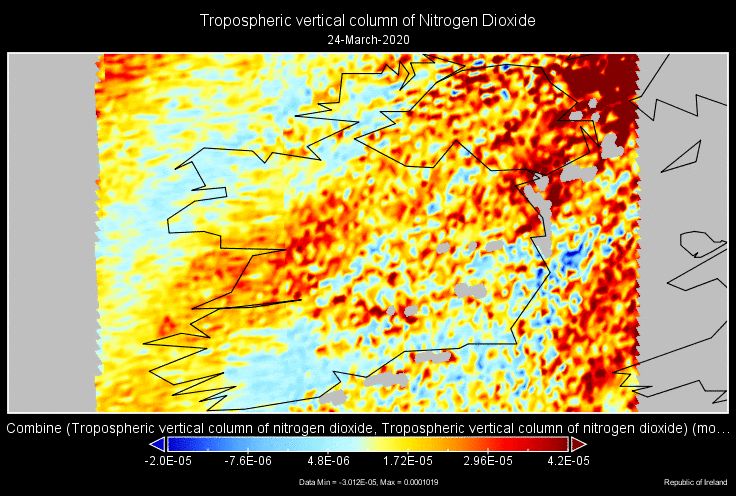}}
\subfloat[25-Mar-2020]{\includegraphics[height=0.2\textwidth]{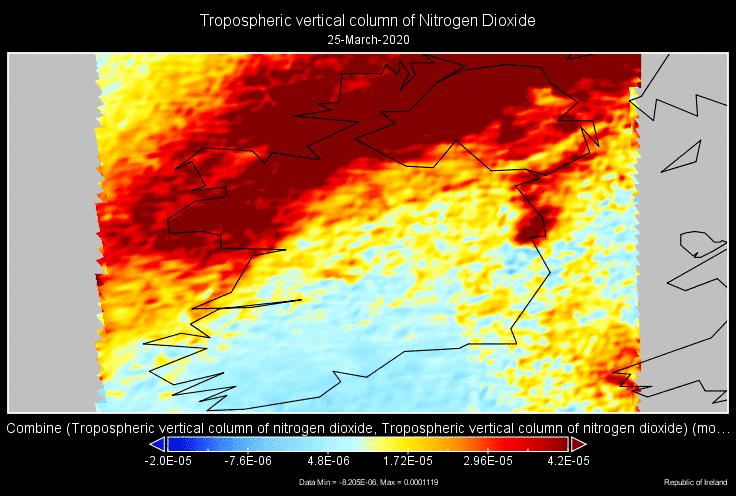}}
\subfloat[26-Mar-2020]{\includegraphics[height=0.2\textwidth]{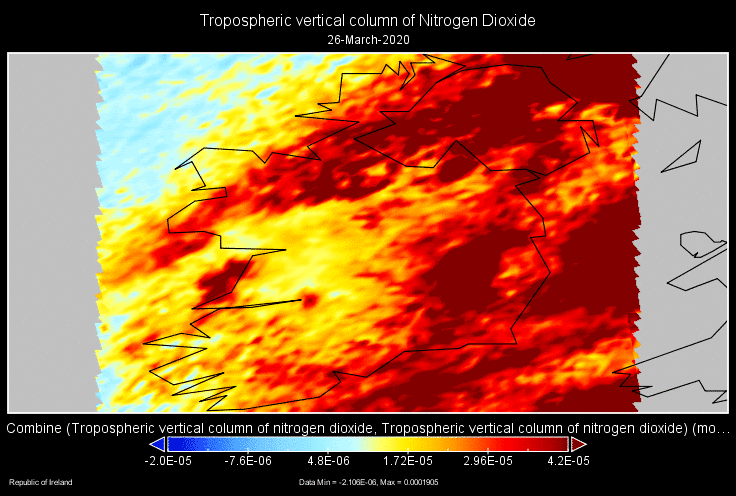}}\\
\subfloat[27-Mar-2020]{\includegraphics[height=0.2\textwidth]{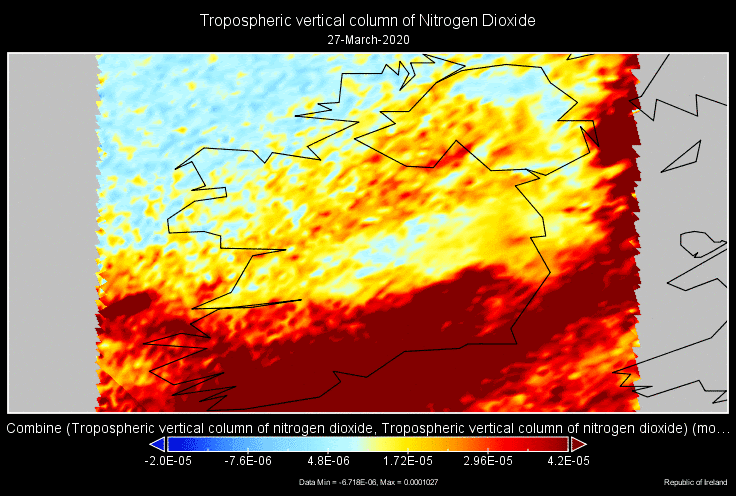}}
\subfloat[28-Mar-2020]{\includegraphics[height=0.2\textwidth]{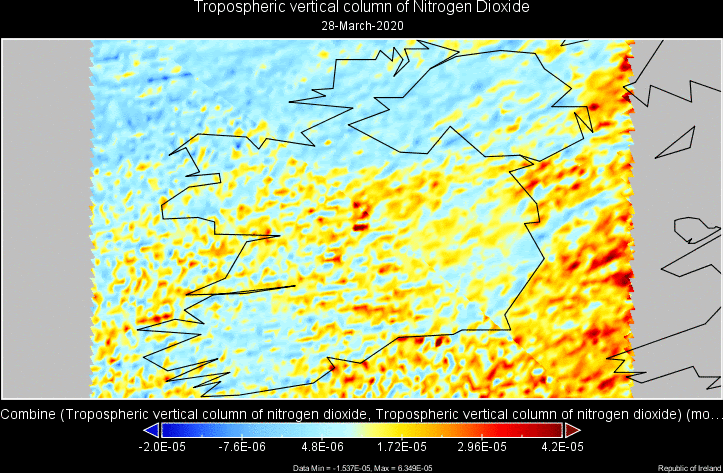}}
\subfloat[29-Mar-2020]{\includegraphics[height=0.2\textwidth]{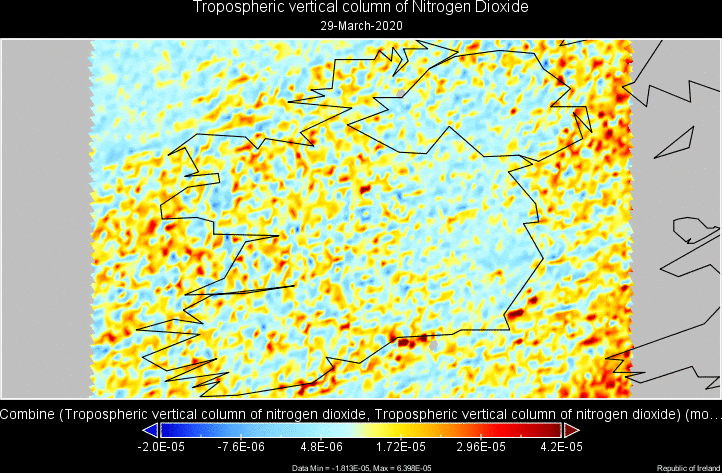}}
\caption{We observe that the concentration of NO$_2$ dropped significantly on 24-March-2020 (day of the lockdown) for Ireland, subsequently all ‘non-essential’ travel were banned after 3 days. We noticed a shape drop again on 28-March-2020 and we observe a clear reduction in NO$_2$ concentration (\textit{cf.} e, f submaps).}
\label{fig:nindia-no2}
\end{figure*}

\begin{figure*}[htb]
\centering
\subfloat[\centering Concentration of NO$_2$ in 2019.]{
\includegraphics[width=0.85\textwidth]{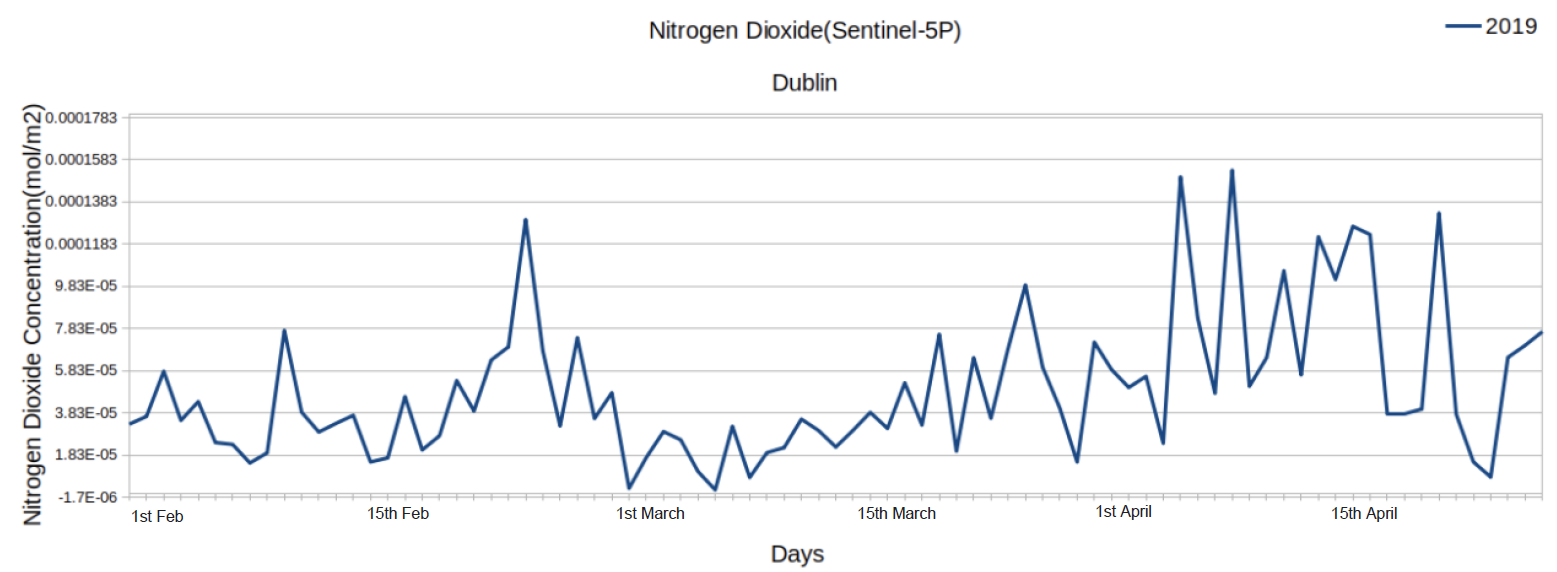}}\\
\subfloat[\centering Concentration of NO$_2$ in 2020.]{
\includegraphics[width=0.85\textwidth]{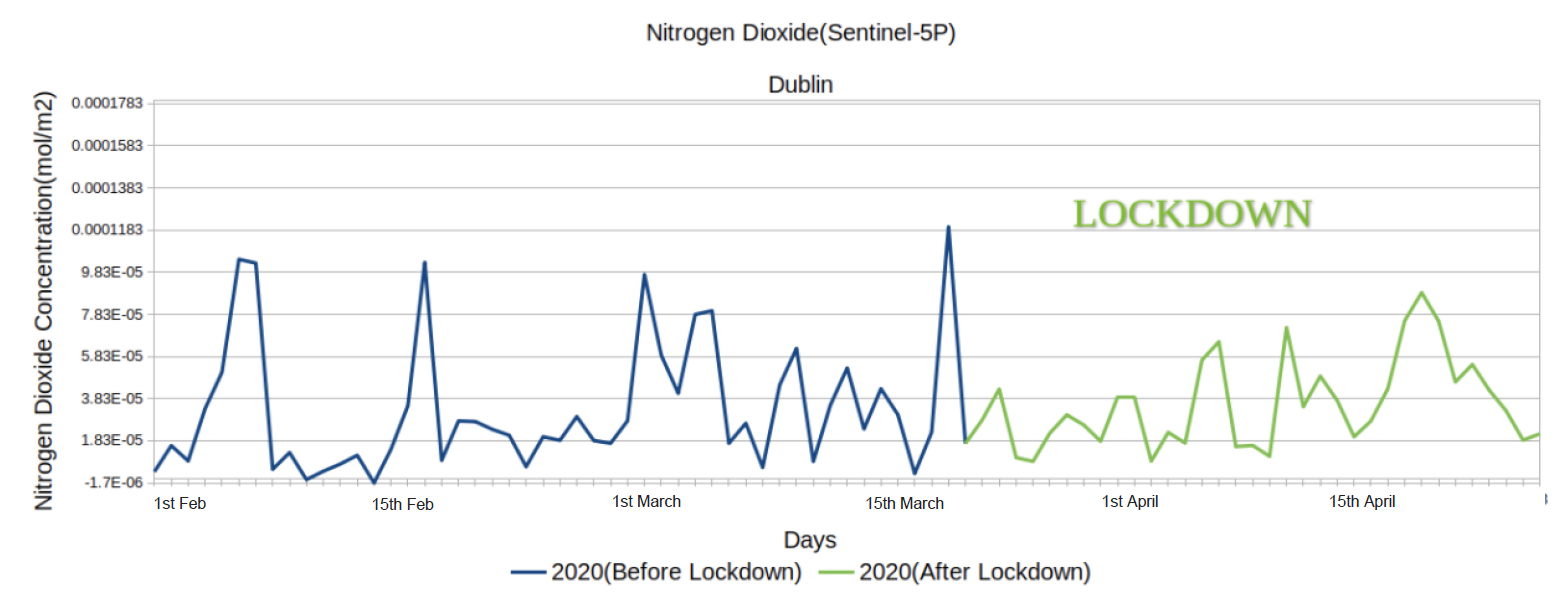}}
\caption{Plot of average daily concentration of nitrogen dioxide obtained from Sentinel-5P for the years (a) 2019, and (b) 2020. We observe that the lockdown significantly reduced the concentration of nitrogen dioxide for the corresponding months of the two years. We show the lockdown period in green color.}
\label{fig:no2-2019-2020}
\end{figure*}

\section{Introduction}
The increasing urbanization of the major cities in the world has put a major impact of the environment. This lead to increased vehicular emissions and related green-house gas emissions to the world. The degradation of the air quality caused serious long-term damage to the lungs, heart disease~\cite{nwosu2019predicting}, and other respiratory diseases~\cite{guarnieri2014outdoor}. Therefore, remote sensing analysts continually recorded the amount of atmospheric pollutants. This assisted in generating preventive measures and providing early health warnings to the citizens. Satellite measurements are usually popular for these cases because of its wide availability~\cite{manandhar2017correlating}. With the recent onset of the COVID19 pandemic, the researchers found an opportune moment to understand the impact of reduced vehicular movement and human activity on the atmospheric air quality. In this paper, we focus our case study for the city of Dublin in Ireland. Ireland confirmed its first confirmed case of COVID-19 on 29-Feb-2020 and thereby saw growth in the number of cases continuously in the month of March. The number of cases suddenly rose in the middle of the month of March, and consequently, schools and colleges were closed down to arrest the spread of the pandemic. Subsequently, the government imposed a strict lockdown to further curtail the spread~\cite{mahato2020effect}. The lockdown was imposed on 24-Mar-2020 and subsequently, three days later all non-essential travel were banned to curtain the spread and avoid community transmission as much as possible. This presented a surprisingly great reduction in pollutant levels. We delve further into the concentration of the individual pollutants obtained from the TROPOMI instrument's dataset. We compare the average concentration of such pollutant with that of the previous year data for the month of March and April to quantify the improvement in the air quality\cite{explo}.

The main contributions of the paper include:

\begin{itemize}
    \item we establish the importance of Sentinel-5P satellite data to monitor the atmospheric pollutants;
    \item we demonstrate that the lockdown induced because of COVID19 significantly reduced the pollutant concentration in Ireland;
    \item we also share the source-code of our methodology in the spirit of reproducible research\footnote{The code used to obtain plots and results in this paper is shared at \url{https://github.com/dkaloni/LockdownAnalysis_Ireland}.}.
\end{itemize}

The rest of the paper is arranged as follows. Subsection~\ref{Data Collection} discusses about the Dataset and the TROPOMI instrument of Sentinel-5P data used for our analysis and the data format of files obtained from the sources. Subsection~\ref{Data Preprocessing} discusses with the methodology used to obtain large sized satellite data, resolving it to usable form and the processing techniques followed to obtain daily average concentration. Subsection~\ref{Pollutant Analysis of Ireland} briefly explains the change in nitrogen dioxide concentration for Ireland followed by Subsection~\ref{Pollutant Analysis of Dublin} which explains the drop in concentration of NO$_2$ in the atmosphere of Dublin using satellite data. Finally, Section~\ref{Future Work} concludes the paper and discusses the future work.

\section{Our Approach}

\subsection{Data Collection}
\label{Data Collection}
This study mainly focuses on Ireland and involved processing data specifically for Dublin.  We systematically analyze the data obtained from satellite observations. The data obtained from TROPOMI\footnote{The TROPOMI data can be accessed from \url{http://www.tropomi.eu/}.} are stored in NetCDF format. The NetCDF stands for Network Common Data Form, and is the format of file used to multidimensional data to study atmospheric features. The Sentinel-5P data uses NetCDF files to share and store files in an array-like structure stacked together. It is generally used to share climate data and for related purposes. The NetCDF4 file contains both the data and the metadata for the product.
The data used for this study is  \texttt{nitrogendioxide\_tropospheric\_column} product of the file which gives the total atmospheric NO$_2$ column between the surface and the top of the troposphere, which stands for tropospheric vertical column of nitrogen dioxide~\cite{NO2Sent1}. 

\subsection{Data Preprocessing}
\label{Data Preprocessing}
The data points included in this region of interest are thereby used for the subsequent computation of pollutant concentration.The measurements are stored and archived in individual NetCDF files. Each NetCDF file indicate the concentration of trace gas for the selected day of the study.  We then combined all the obtained files into a single file by adding a new dimension of time to the dataset for comparative study\footnote{The NCO is a set of command line utilities used to operate NetCDF files efficiently, available at \url{http://nco.sourceforge.net/}}. This combined file provides us with data points for the area spread across the time dimension.

Our obtained combined NetCDF file provides us the estimate of the average daily concentration of the various atmospheric pollutants. We process these files to obtain the pollutant concentrations of the different pollutant gases. We replace the \textit{NaN} values of the data grid to $0$, and thereby calculate the sum of the all the measurement values in the particular grid. We divide the resultant sum by the number of non-zero values to have a daily average for the region. This enables us to calculate the average concentration of a particular pollutant over the selected region of Dublin~\cite{zhang2020estimating}.

\subsection{Pollutant Analysis of Ireland}
\label{Pollutant Analysis of Ireland}
We performed the primary analysis on Ireland using \texttt{panoply} (geo-referenced data viewer by NASA Goddard Institute of Space Studies). We generated the spatial concentration maps of nitrogen dioxide for the country from global dataset for dates during lockdown. The nitrogen dioxide concentration was above average on 24-Mar-2020, which dropped significantly on 28-Mar-2020, when strict lockdown was imposed. Subsequent day of 29th March followed the trend of sharp reduction in pollutant concentration. Lockdown stopped nearly all vehicular and industrial activities around the country, which was depicted by improvement in the NO$_2$ pollutant concentration plots. Figure~\ref{fig:nindia-no2} shows the daily NO$_2$ concentration plot for Ireland for initial days of lockdown, and we observe a sharp drop in the NO$_2$ concentration because of the lockdown.

\subsection{Pollutant Analysis of Dublin}
\label{Pollutant Analysis of Dublin}

To subset Dublin from the global dataset, latitude range from 53.24$^{\circ}$ to 53.41$^{\circ}$ and longitude range from -6.11$^{\circ}$ to -6.45$^{\circ}$ was fixed and the concentration data points for the area were chosen. The combined file provided us with data points on the concentration of nitrogen dioxide for the area spread across the time dimension for better analysis. Daily average concentration plot for Feb-2019, Mar-2019 and Apr-2019 is shown in Figure~\ref{fig:no2-2019-2020}(a) and daily average concentration plot for Feb-2020, Mar-2020 and Apr-2020 are shown in Fig.~\ref{fig:no2-2019-2020}(b).
The last 10 days of March 2019 recorded the average NO$_2$ concentration of $5.489$ $\mu$mol/m$^2$, whereas the last 10 days of Mar-2020 recorded the average concentration of $3.044$ $\mu$mol/m$^2$, a drop of about 44.54\%.

The average concentration for the initial 10 days of Apr-2020 recorded $3.072$ $\mu$mol/m$^2$, which indicated towards the similar trend of drop in NO$_2$ concentration post lockdown. Figure~\ref{fig:no2-2019-2020}(b) shows the trend of continuous concentration drop after lockdown. The calculated average concentration of NO$_2$ for Apr-2019 was $7.075$ $\mu$mol/m$^2$ whereas for Apr-2020 it was $3.831$ $\mu$mol/m$^2$, recording a significant drop of 45.8\%. The significant improvement in NO$_2$ concentration was because of the strict lockdown imposed in Dublin for the month of April~\cite{dewanshreport}. The time-series plot for 2019 and 2020 shows the drop in NO$_2$ concentration post 24-Mar-2020 (lockdown was imposed) in comparison to the same period in Mar-2019.

\begin{figure}[htb]
\begin{center}
\includegraphics[width=0.5\textwidth]{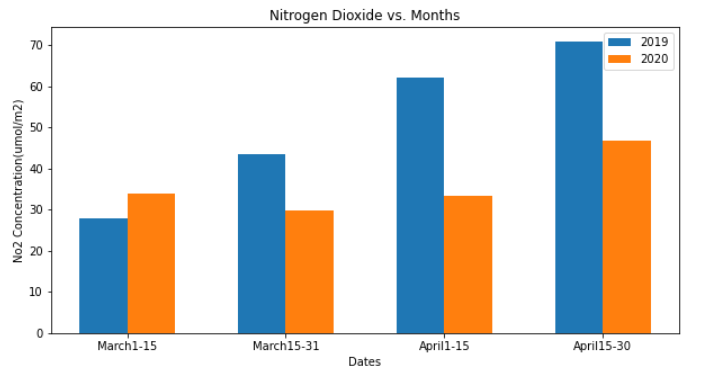}
\caption{We compute the fifteen-days average nitrogen dioxide concentration for the months of March and April, and demonstrate it via a stacked bar chart for both the years -- 2019 and 2020. We observe from the bar plot that there is a reduced NO$_2$ concentration from second half of March 2020 and beyond.
\label{fig:2019-2020}}
\end{center}
\end{figure}

Figure~\ref{fig:2019-2020} shows the average fifteen days concentration comparison for both the years of the city for March and April. The city noticed a increasing trend of NO$_2$ concentration for Mar-Apr 2019, the trend for 2020 was very much different due to lockdown. The second half of Mar-2020 recorded a 31.17\% drop as compared to same period last year which was mainly due to the partial lockdown in the city. A complete lockdown further improved the NO$_2$ concentration for the first half of Apr-2020 by 46.35\% as compared to the first half of Apr-2019.

\section{Conclusions \& Future Work}
\label{Future Work}
Such remote sensing techniques of analysis can be beneficial for monitoring the atmosphere in remote regions of the earth which lack ground-based monitoring stations with zero new installments. The COVID-19 lockdown brought a rare opportunity for researchers to understand the environment much better and thereby attempt to influence the policymakers to implement stricter environmental laws. Our future work include the short-term forecasting of pollutant data from historical data. We intend to use the popular LSTM-based models~\cite{jain2020forecasting} to learn the underlying pattern of such time-series pollutant data and propose relevant recommendations to city planners and environmental agencies. Furthermore, we also plan to use a multi-sensor framework involving ground-based sky cameras~\cite{jain2021wsi} and monitoring stations~\cite{manandhar2018systematic}, in addition to the satellite measurements.


\end{document}